\documentclass[pre,a4paper,superscriptaddress,twocolumn,showpacs,amsmath,amssymb,floatfix]{revtex4}

\pdfoutput=1
\usepackage{graphicx}
\usepackage{amssymb}
\usepackage{float}

\usepackage{color}

\begin{document}

\title{Small world of Ulam networks for chaotic Hamiltonian dynamics}

\author{Klaus M.~Frahm}
\affiliation{\mbox{Laboratoire de Physique Th\'eorique, IRSAMC, 
Universit\'e de Toulouse, CNRS, UPS, 31062 Toulouse, France}}
\author{Dima L.~Shepelyansky}
\affiliation{\mbox{Laboratoire de Physique Th\'eorique, IRSAMC, 
Universit\'e de Toulouse, CNRS, UPS, 31062 Toulouse, France}}

\date{July 13, 2018}

\begin{abstract}
We show that the Ulam method applied to dynamical symplectic maps
generates Ulam networks which belong to the class of small world networks
appearing for social networks of people, actors, power grids, 
biological networks and Facebook.
We analyze the small world properties of Ulam networks
on examples of the Chirikov standard map and the Arnold cat map
showing that the number of degrees of separation, or the Erd\"os number,
grows logarithmically with the network size for the regime of strong chaos.
This growth is related to the Lyapunov instability of chaotic dynamics.
The presence of stability islands leads to an algebraic growth
of the Erd\"os number with the network size.
We also compare the time scales related with the Erd\"os number
and the relaxation times of the Perron-Frobenius operator
showing that they have a different behavior. 
\end{abstract}



%

\maketitle

\section{Introduction} 
\label{sec1}
In 1960 Ulam proposed a method \cite{ulam},
known now as the Ulam method, for generating discrete, a finite cell 
approximate 
of the Perron-Frobenius operator for a chaotic map in a continuous phase space.
The transition probabilities from one cell to others
are determined from an ensemble of trajectories
which generates the probabilities of Markov transitions \cite{markov}
between cells after one map iteration. In this way the Ulam method
produces the Ulam networks with weighted probability transitions between
nodes corresponding to phase space cells.  
For one-dimensional (1D) fully chaotic maps \cite{lichtenberg,cvitanovic}
the convergence of the discrete dynamics of 
this Ulam approximate of the 
Perron-Frobenius operator (UPFO), at cell size going to zero, 
to the continuous dynamics has been mathematically proven in
\cite{li}. The properties of UPFO was studied for 1D 
\cite{tel,kaufmann,froyland2007}
and 2D \cite{ding,liverani,froyland2008a,froyland2008b} maps.
It was shown that the UPFO finds useful applications for analysis of 
dynamics of molecular systems \cite{schutte}
and coherent structures in dynamical flows \cite{froyland2009physd}.
Recent studies \cite{zhirov,ermannweyl}
demonstrated similarities between the UPFO,
corresponding to Ulam networks, 
and the Google matrix of complex directed networks 
of World Wide Web, Wikipedia, world trade and other systems
\cite{brin,meyerbook,rmp2015}. 

From a physical point of view 
the finite cell size of UPFO corresponds to the introduction
of a finite noise with amplitude given by a 
discretization cell size. For dynamical maps
with a divided phase space, like the Chirikov standard map
\cite{chirikov}, such a noise leads to the destruction
of invariant Kolmogorov-Arnold-Moser (KAM) curves 
\cite{lichtenberg,arnold,chirikov} so that 
the original Ulam method is not operating in a correct way
for such maps. However, the method can be considered in its
generalized form \cite{frahmulam} 
when the Markov transitions are generated 
by specific trajectories starting only inside one chaotic component
thus producing Markov transitions between cells
belonging only to this chaotic component. Due to ergodicity on the
chaotic component even only one long chaotic trajectory can 
generate a complete UPFO avoiding the destruction of KAM curves
and stability islands. It was also shown
numerically that the spectrum of the finite size UPFO matrix
converges to a limiting density at cell size going to zero \cite{frahmulam}.

Certain similarities between the spectrum of UPFO matrices of Ulam networks
and those of Google matrix of complex directed networks have already been 
discussed in the literature (see e.g. \cite{rmp2015}).
Here, we address another feature of Ulam networks 
showing that they have small world properties meaning 
that almost any two nodes are indirectly connected by a 
small number of links. 
Such a small world property with its {\it six degrees of separation}
is typical for social networks of people \cite{milgram},
actors, power grids, biological and 
other networks \cite{strogatz,newman,dorogovtsev}.
Thus the whole Facebook network of about 700 million users
has only four degrees of separation \cite{vigna2012}.

The paper is organized as follows:
Section \ref{sec2} presents the main properties of the
two symplectic maps considered,
construction of Ulam networks is described in Section \ref{sec3},
small world properties of Ulam networks are analyzed in Section \ref{sec4},
the relaxation rates of the coarse-grained Perron-Frobenius operator
are considered in Section \ref{sec5}. The obtained results are discussed
in the last Section \ref{sec6}.

\section{Dynamical symplectic maps}
\label{sec2}
We analyze the properties of Ulam networks for two examples being 
the Chirikov standard map \cite{chirikov} and the Arnold cat map \cite{arnold}.
Both maps capture the important generic features of Hamiltonian dynamics and 
find a variety of applications for the description of real physical systems
(see e.g. \cite{stmapscholar}).

The Chirikov standard map has the form:
\begin{equation}
\label{eq_stmap}
{\bar p} = p + \frac{K}{2\pi} \sin (2\pi x) \; , \;\; 
{\bar x} = x + {\bar p} \;\; ({\rm mod} \; 1) \;.
\end{equation}
Here bars mark the variables after one map iteration
and we consider the dynamics to be periodic on  a torus so that
$0 \leq x \leq 1$, $0 \leq p \leq 1$. It is argued that the last
KAM curve is the one with the golden rotation number being destroyed 
at critical $K_c=K_g=0.971635406...$ \cite{mackay}.
Indeed, further mathematical analysis \cite{percival}
showed that all KAM curves are destroyed
for $K \ge 63/64$ while the numerical analysis
 \cite{chirikov2000} showed that  
$K_c-K_g  < 2.5 \times 10^{4}$. Thus
it is most probable that $K_c=K_g$ and the golden KAM curve is
the last to be destroyed (see also the review \cite{meiss}).

The Arnold cat map \cite{arnold} of the form, 
\begin{equation}
\label{cateq1}
\bar{p}=p+x \; \mbox{(mod} \;\mbox{L)}\;\;, \;\; 
\bar{x}=x+\bar{p} \;\mbox{(mod} 
\;\mbox{1)}\;,
\end{equation}
is the cornerstone model of classical dynamical chaos 
\cite{lichtenberg}. 
This symplectic map belongs to the class of Anosov systems,
it has the positive Kolmogorov-Sinai entropy
$h=\ln[(3+\sqrt{5})/2)]\approx 0.9624$
and is fully chaotic \cite{lichtenberg}. 
Here the first equation can be seen as a kick which changes the momentum 
$p$ of a particle on a torus 
while the second one corresponds to a free phase rotation
in the interval $-0.5\leq x < 0.5$; bars mark the new values of 
canonical variables $(x,p)$.
The map dynamics takes place on a torus of integer length $L$ in the 
$p$-direction with $-L/2 < p \leq L/2$. 
The usual case of the Arnold cat map
corresponds to $L=1$ but it is more interesting to study 
the map on a torus of longer integer size 
$L>1$ generating a diffusive dynamics in $p$ \cite{demon,arnoldcat}.
For $L \gg 1$ the diffusive process for the probability density $w(p,t)$
is described by the Fokker-Planck  equation:
\begin{equation}
\label{cateq2}
\frac{\partial w(p,t)}{\partial t} = \frac{D}{2}
\frac{\partial^2 w(p,t)}{\partial p^2}\ ,
\end{equation}
with the diffusion coefficient $D \approx \langle x^2\rangle= 1/12$ 
and $t$ being iteration time. As a result for times
$t \gg L^2/D$ the distribution converges to the ergodic equilibrium
with a homogeneous density in the plane $(x,p)$ \cite{arnoldcat}.

\section{Construction of Ulam networks}
\label{sec3}
We construct 
the Ulam network and related UPFO for the map (\ref{eq_stmap}) 
as described in \cite{frahmulam}. First we reduce the phase space 
to the region $0\le x<1$ and $0\le p< 0.5$ exploiting the symmetry $x\to 1-x$ 
and $p\to 1-p$. The reduced phase space is divided into $M\times (M/2)$ cells 
with certain integer values $M$ in the range $25\le M\le 3200$. To determine 
the classical transition probabilities between cells we 
iterate one very long trajectory of $10^{12}$ iterations starting inside 
the chaotic component at $x=p=0.1/(2\pi)$ and count the number of transitions 
from a cell $i$ to a cell $j$. 
Depending on the value of $K$ it is possible that 
there are stable islands or other non-accessible regions where the 
trajectory  never enters. This corresponds to certain cells that 
do not contribute to the Ulam network. In practice, we perform trajectory 
iterations only for the 
largest two values $M=3200$, $M=2240$ and apply an exact renormalization 
scheme to reduce successively the value of $M$ by a factor of 2 down to 
$M=25$ and $M=35$ (for these two cases the vertical cell-number is chosen as 
$(M+1)/2$ with the top line of cells only covering half cells). 
We consider the dynamics for four different values of $K$: 
the golden critical value $K=K_g=0.971635406$, $K=5$, $K=7$ and $K=7+2\pi$.
There are small stability islands for the last three cases. 
The original Ulam method \cite{ulam} 
computes the transition probabilities from one cell 
to other cells using many random initial conditions per cell but for the 
Chirikov standard map this would imply that the implicit coarse graining of 
the method produces a diffusion into the stable islands or other classically 
non-accessible regions which we want to avoid. The typical network size 
(of contributing nodes/cells) is approximately $N_d\approx M^2/2$ 
($N_d\approx M^2/4$) for the cases with $K\ge 5$ ($K=K_g$). 

For the Arnold cat map (\ref{cateq1}) we divide the phase space 
$-0.5\le x<0.5$ and $-L/2\le p<L/2$ into $M\times LM$ cells where in this work 
we mostly choose $L=3$ and $M$ is taken from a sequence of prime numbers 
starting with $M=29$ and increasing $M$ roughly by a factor of $1.4$ in order 
to minimize certain arithmetic effects from non-prime numbers. Since the 
Arnold cat map does not have any inaccessible regions, both variants of the 
Ulam method, with many random initial conditions or one long trajectory 
(using a suitable irrational choice of the initial position) work very well. 

However, due to the exact linear form of (\ref{cateq1}) it is even possible 
to compute directly very {\em efficiently} and {\em exactly} 
(without any averaging procedure) the transition probabilities. Details of 
this procedure for the Arnold cat map together with a discussion of related 
properties of the UPFO for the standard map are given in Appendix 
\ref{appenda}. The results for the UPFO for the cat map given 
below in this work have all been obtained for the exact UPFO computed 
in this way.

\section{Small world properties of Ulam networks}
\label{sec4}
To study the small world properties 
of the Ulam networks we compute a quantity which we call the 
Erd\"os number $N_E$ (or number of degrees of separation) 
\cite{erdos,dorogovtsev}. 
This number represents the minimal number of links 
necessary to reach indirectly a specific node via other intermediate 
nodes from a particular node called the hub. Here the (non-vanishing) 
transition probabilities are not important and only the existence of a link 
between two nodes is relevant. The recursive computation of $N_E$ for all 
nodes can be done very efficiently for large networks by keeping a list 
of nodes with same $N_E$ found in the last iteration and which 
is used to construct a new list of nodes with $N_E$ increased by 
unity as all nodes being connected to a node of the initial list and 
not yet having a valid smaller value of $N_E$ (for nodes found in a 
former iteration). After each iteration the list will be updated with the new 
list and the initial list of this procedure at $N_E=0$ is chosen to contain 
one node being the hub. 

\begin{figure}[t]
\begin{center}
\includegraphics[width=0.48\textwidth]{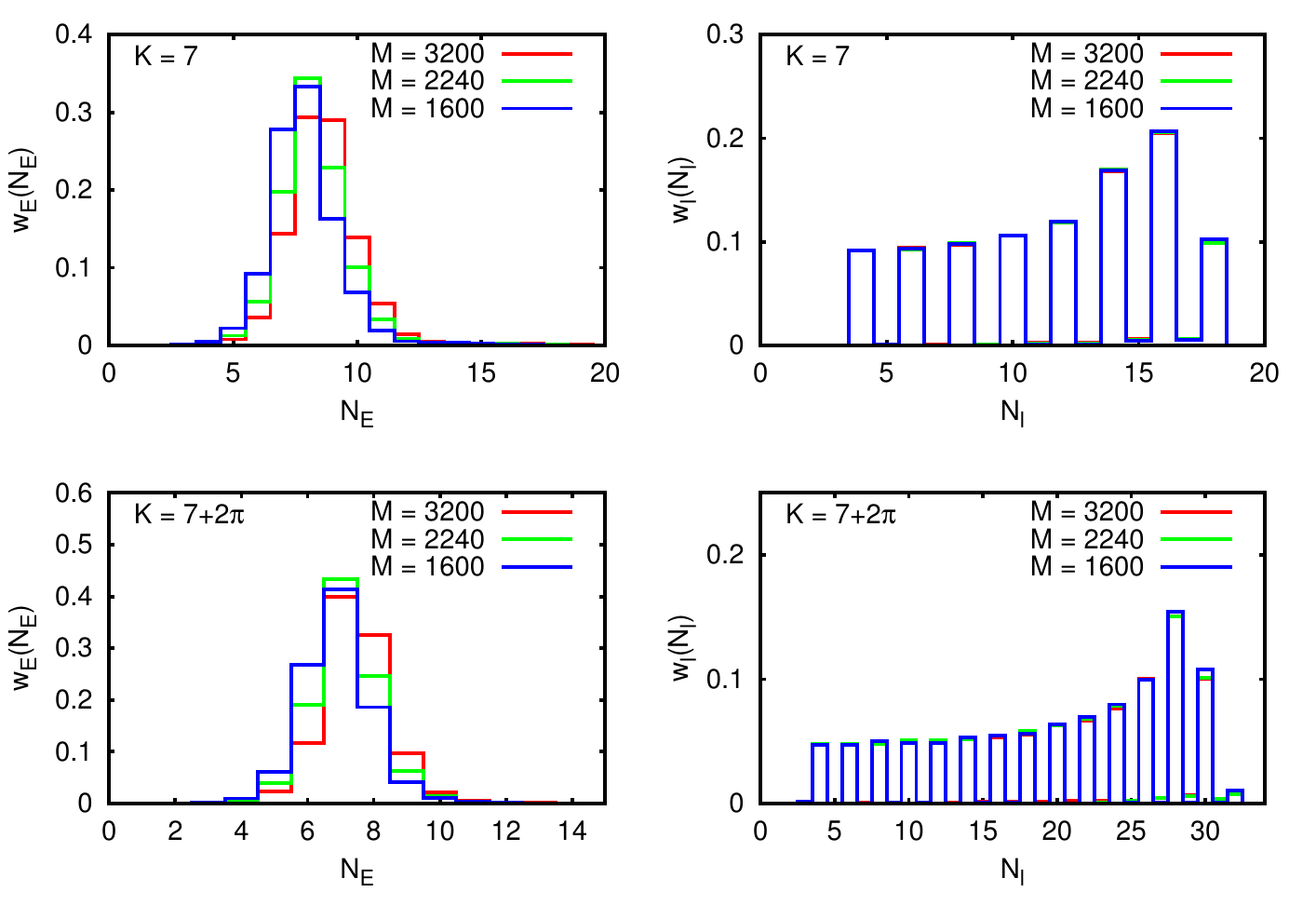}
\end{center}
\caption{\label{fig1}
Left panels: probability distribution $w_E$
of Erd\"os number $N_E$ for the Ulam network
of the Chirikov standard map at $K=7$ (top panel)
and $K=7+2\pi$ (bottom panel) for 
three different numbers of nodes (cells) $N_d\approx M^2/2$ 
using a hub cell at $x=0.1/(2\pi), p=0$. 
Right panels: probability distribution
$w_l$ of number of links $N_l$ per node of the networks of left panels. 
In all cases the Ulam network was constructed with one trajectory of 
$10^{12}$ iterations using the initial condition 
$x=0.1/(2\pi)$, $p=0.1/(2\pi)$.
}
\end{figure}

Fig.~\ref{fig1} shows the probability distributions $w_E(N_E)$ of the 
Erd\"os number $N_E$ (using a hub cell at $x=0.1/(2\pi)$, $p=0$) 
and the number of links $N_l$ per node of the UPFO for the Chirikov standard 
map at $K=7$ and $K=7+2\pi$ for the three 
largest values of $M=1600,\,2240,\,3200$  considered. 
The distributions of $N_E$ are quite sharp with mean values of $N_E$ 
around 8 (or 7) for $K=7$ ($K=7+2\pi$) being slightly increasing with $M$. 
Even though the maximal possible values are larger 
($N_E^{(\rm max)}=33$ for $K=7$ and $N_E^{(\rm max)}=21$ for $K=7+2\pi$ at
$M=3200$) the big majority of nodes have a value $N_E\le 12$ 
($10$) for $K=7$ ($K=7+2\pi$) clearly confirming the small world structure 
of these networks. 

The corresponding distribution $w_l(N_l)$ of number of links $N_l$ 
shows that $N_l$ takes essentially 
only even values in the range $4\le N_l\le N_l^{(\rm max)}$ with 
$N_l^{(\rm max)}=18$ ($32$) for $K=7$ ($K=7+2\pi$). This behavior can be 
understood in the framework of the discussion in Appendix \ref{appenda} 
showing that the image of an initial 
square cell is (up to non-linear corrections) a parallelogram with extreme 
points (relative to a certain reference cell) $\Delta s(\xi_0,\xi_0)$ and 
$\Delta s(\xi_0+A+2,\xi_0+A+1)$ where $\xi_0$ is a quasi-random uniformly 
distributed quantity in the interval $\xi_0\in[0,1[$ and 
$\Delta s=1/M$ is the linear cell size. Here we 
assume that $A=K\cos(2\pi \Delta s\,x_i)>0$ (the argumentation 
for $A<0$ is rather similar with $A\to|A|$). The parallelogram 
covers in horizontal direction nearly always two cells and in diagonal 
direction $\lceil \xi_0+A+1\rceil\ge 2$ cells where $\lceil u\rceil$ 
is the ceil function of $u$ being the smallest integer larger or equal 
than $u$.
Therefore typical values of $N_l=2\lceil \xi_0+A+1\rceil$ are 
indeed even numbers with $4\le N_l\le N_l^{(\rm max)}$ where 
$N_l^{(\rm max)}=2\lceil 2+K\rceil$ is in agreement with the 
observed values in Fig.~\ref{fig1}. 

Actually for 
$K=7+2\pi\approx 13.283$ we also understand that the probability 
for $N_l=N_l^{(\rm max)}=32$ is quite strongly reduced because 
even for maximal $A=K$ we need that the offset satisfies $\xi_0>1-0.283$ which 
is statistically less likely. Apart from this there is also a slight 
increase of histogram bins with larger values of $N_l$ due to the 
cosine factor in $A$ applied on a uniformly distributed phase. 
For sufficiently large $M$ this 
argumentation does not depend on system/network 
size. We mention that for very small values of $M$ there are deviations from 
this general picture, with some small probabilities for odd values of 
$N_l$ due to boundary effects, also related to stable islands and 
inaccessible phase space regions (especially for $K=K_g$). For the 
largest values $M=3200,\,2240$ and $K=7+2\pi$ 
the figure shows some small deviations due to statistical fluctuations since the 
average ratio of trajectory transitions per link 
$10^{12}/(N_d N_l^{(\rm max)})\approx 6000$ is rather modest. 
Furthermore, the data for $K=5$ (not shown in Fig.~\ref{fig1}) are also in 
agreement with this general picture with $N_l^{(\rm max)}=14$ and 
typically $N_E\approx 11\pm 3$.

\begin{figure}[t]
\begin{center}
\includegraphics[width=0.48\textwidth]{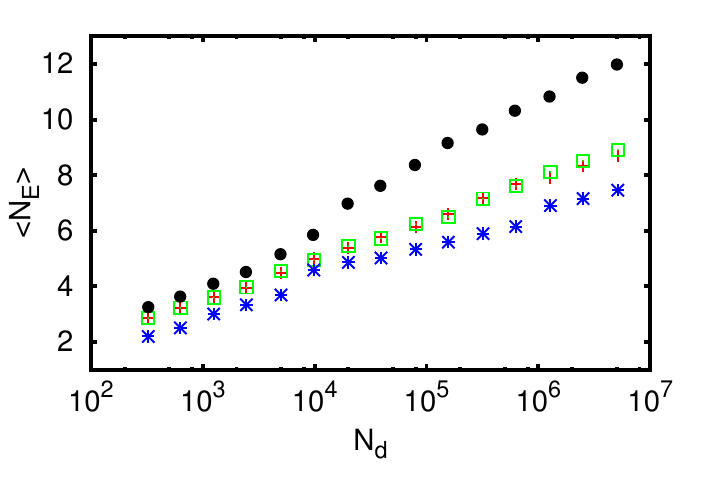}
\end{center}
\caption{\label{fig2}
Dependence of the average  Erd\"os number $\langle N_E\rangle $
on the number of nodes (cells) $N_d$ for the Ulam network
of the Chirikov standard map at $K=5$ (top data points, black dots), 
$K=7$ (both sets of center data symbols of red pluses and green squares) 
and $K=7+2\pi\approx 13.28$ (bottom data points, blue stars). 
The hub cell is either at $x=0.1/(2\pi)$, $p=0$ (top, center with 
red plus symbols and bottom) or at $x=0.1/(2\pi)$, $p=0.1/(2\pi)$ 
(center data with green squares). }
\end{figure}

According to Fig.~\ref{fig2} the average Erd\"os number for the three cases 
with $K\ge 5$ behaves approximately as 
\begin{equation}
\label{eq_erdos}
\langle N_E\rangle\approx C_1+C_2\ln (N_d) \; .
\end{equation}
Here $C_1$, $C_2$ are some numerical constants which have no significant dependence on the hub choice 
as long it is not close to some stable island 
or similar. The typical values of $C_2$ are close to $h^{-1}$ with 
$h=\ln(K/2)$ being the Lyapunov exponent of the standard map (for $K > 4$) \cite{chirikov}. 
This is due to the theoretically expected behavior $N_f(N_E)\approx e^{hN_E}$ 
for $N_E<\langle N_E\rangle$ and where $N_f(N_E)=N_d w_E(N_E)$ 
is the number of cells 
indirectly connected to the hub after $N_E$ iterations. This theoretical 
behavior is rather well confirmed by the data of left panel of Fig.~\ref{fig1} 
(when presented in log presentation for the $y$-axis and multiplied 
with $N_d$). The exponential increase saturates at $N_E=\langle N_E\rangle$ 
with $e^{h\langle N_E\rangle}\approx \alpha N_d$ and $\alpha$ being a 
constant of order of unity implying $C_2=1/h$ and $C_1=\ln(\alpha)/h$. 

\begin{figure}[t]
\begin{center}
\includegraphics[width=0.48\textwidth]{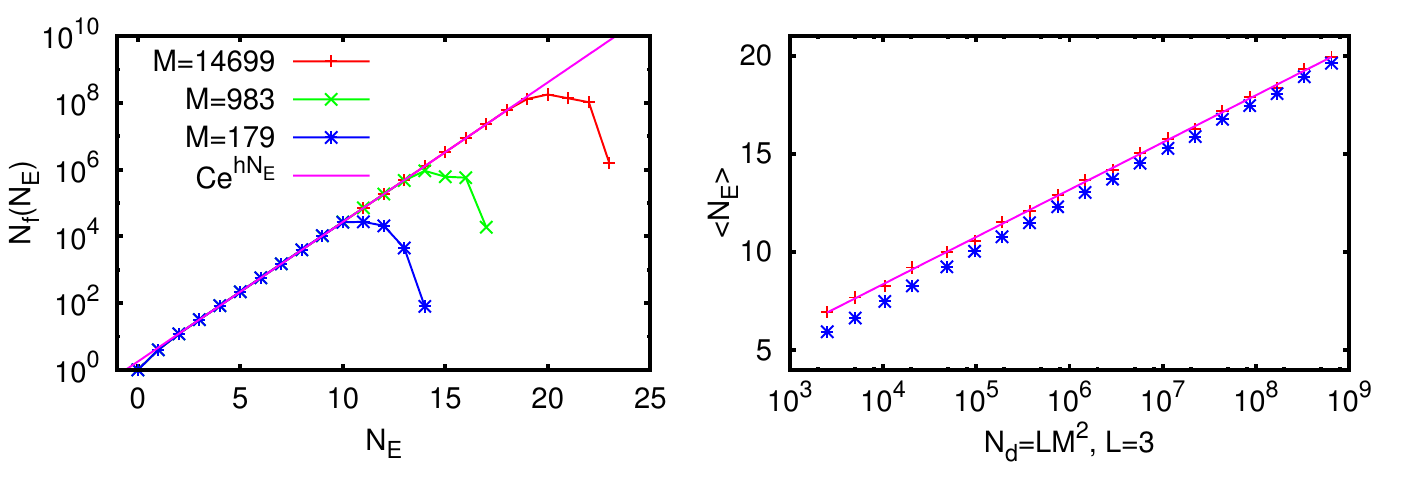}
\end{center}
\caption{\label{fig3}
Properties of the Erd\"os number $N_E$ for the Ulam network
of the Arnold cat map on a torus with $L=3$ with hub cell at 
position $x=p=0$. 
Left panel: frequency distribution $N_f(N_E)$ for 
three different values of $M$. The pink line corresponds to the 
fit $N_f(E)=C\,e^{h N_E}$ for $M=14699$ using the fit range 
$2\le N_E\le 18$ with $C=1.777\pm 0.005$ and 
$h=0.9629\pm 0.0002$. 
Right panel: dependence of average $\langle N_E\rangle$ 
on the number of nodes (cells) $N_d=LM^2$ (red plus symbols). 
The blue stars correspond to a restricted average over nodes being in the 
center square box with $|x|< 0.5$ and $|p|<0.5$ (instead of $|x|<0.5$ and 
$|p|<L/2$ for the full average). The pink line corresponds to the fit
(of top data points) $\langle N_E\rangle =d+\ln(N_d)/\tilde h$ with 
$d=-1.25\pm 0.08$ and $\tilde h=0.957\pm 0.005$. 
The two values of $h$ and $\tilde h$ compare to the theoretical 
Lyapunov exponent $h_{\rm th}=\ln[(3+\sqrt{5})/2]\approx 0.9624$. 
The Ulam network for the 
Arnold cat map was constructed from exact theoretical transition 
probabilities as described in Appendix \ref{appenda}. The number of links 
per node is constant for all nodes: $N_l=5$ ($4$ or $6$) 
if $M$ and $LM$ are odd ($M$ and $LM$ even or $M$ odd and $LM$ even 
respectively).
}
\end{figure}

We have performed a similar analysis of $N_l$ and 
$N_E$ also for the Arnold cat map. Here the link number $N_l$ is constant 
for all nodes with values $4$, $5$ or $6$ depending on the 
parity of $M$ or $LM$ as explained in Appendix \ref{appenda}. 
The behavior 
of $N_E$ is presented in Fig.~\ref{fig3} showing the frequency distribution 
$N_f(N_E)=N_d w_E(N_E)$ (left panel) and the dependence $\langle N_E\rangle$ 
on $N_d$ (right panel) for $L=3$ and several (prime) values of $M$. 
The expected theoretical behavior of both quantities is very clearly confirmed 
providing accurate fit values of the Lyapunov exponent being numerically 
very close to the theoretical value 
$h_{\rm th}=\ln[(3+\sqrt{5})/2]\approx 0.9624$. 
Furthermore, the saturation of the exponential growth of $N_f(N_E)$ 
for $N_E\ge\langle N_E\rangle$ happens quite abruptly with 
$N_E^{(\rm max)}=\langle N_E\rangle+3$. 
We also computed the restricted average of $N_E$ over the center square box 
(out of $L=3$ squares) with $|x|< 0.5$ and $|p|<0.5$ which turns out 
to be quite close to the full average with $|x|< 0.5$ and $|p|<L/2$ 
showing that for the Erd\"os number the diffusive dynamics is apparently 
not very relevant. 

\begin{figure}[t]
\begin{center}
\includegraphics[width=0.48\textwidth]{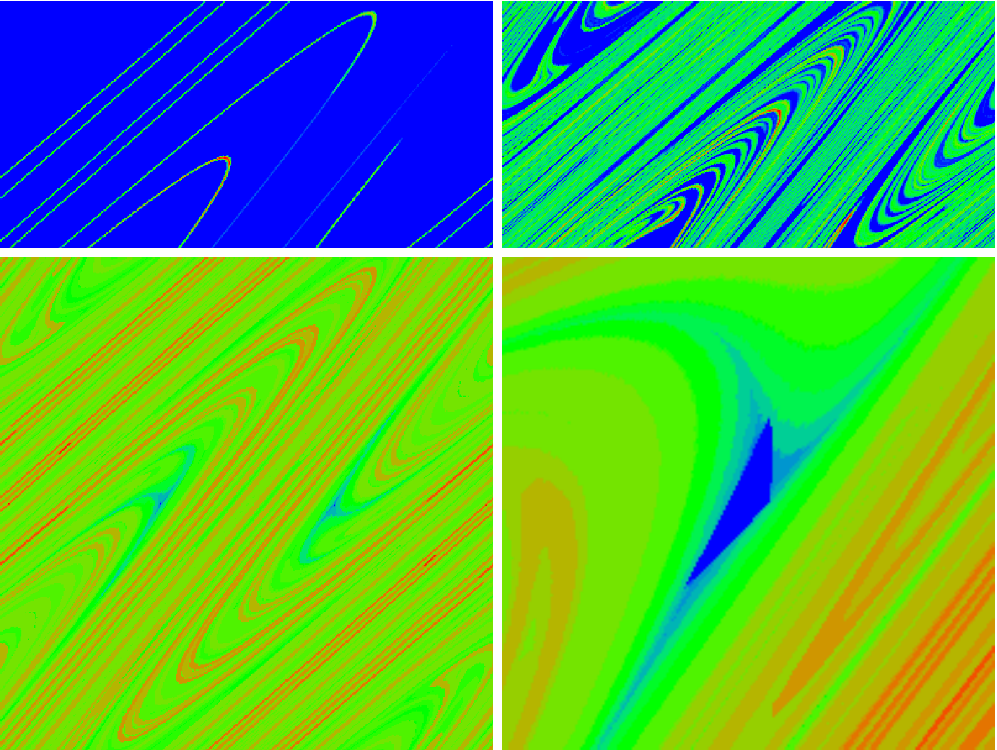}
\end{center}
\caption{\label{fig4}
Top: color density plot of the probability distribution
in the phase plane $(x,p)$ obtained 
after $t=4$ (left panel) or $t=7$ (right panel) iterations of the UPFO 
($M=400$) for the Chirikov standard map at $K=7$ with 
an initial state being localized in one cell at $x=p=0.1/(2\pi)$.
The colors red/green/blue correspond to maximum/medium/minimal values. 
Bottom: density plots of the Erd\"os number of nodes/cells for the 
Ulam network (standard map, $K=7$, same hub position as in Fig.~\ref{fig1}) 
in the phase plane $(x,p)$ with red/green/light blue corresponding 
to smallest/medium/largest Erd\"os numbers. The dark blue cells correspond 
to non-accessible islands which do not contribute as nodes for the 
Ulam network. Left panel corresponds to one full square box of the phase space 
given by $0\le x<1$ and $-0.5\le p<0.5$ for $M=400$. Right panel 
shows a zoom of $200\times 200$ cells with bottom left corner at 
cell position $(935,1500)$ for $M=3200$ and containing the left of 
the two islands (for $K=7$). In both bottom panels data for cells with 
$p<0$ are obtained from the symmetry: $p\to -p$ and $x\to 1-x$. 
}
\end{figure}

To understand the spatial structure of the Erd\"os number of nodes we 
show in top panels of Fig.~\ref{fig4} density plots of the phase space 
probability distribution after a few iterations of the UPFO for the 
map (\ref{eq_stmap}) at $K=7$ and $M=400$ applied to an initial cell state. One 
can clearly identify the chaotic spreading along a one-dimensional manifold 
which fills up the phase space due to refolding induced by the periodic 
boundary conditions. The lower panels show the full spatial 
distribution of the Erd\"os number by a color plot 
using red (green, light blue) for nodes with smallest (medium, largest) 
Erd\"os number. Dark blue is used for non-accessible nodes 
in the stable islands which have no Erd\"os number. Furthermore, 
for a better visibility we consider a full square box 
with $0\le x<1$ and $-0.5\le p<0.5$ where the 
data for $p<0$ is obtained by the transformation $x\to 1-x$ and $p\to -p$ 
from the data with $p\ge 0$. In this way the two small stable islands at 
$p=0$ for $K=7$ have a full visibility 
(the influence of orbits sticking near these islands 
on Poincar\'e recurrences is discussed in \cite{chirikov2000k7}). 
Nodes with the smallest Erd\"os number 
follow the same one-dimensional unstable manifold as the 
chaotic stretching and nodes with maximal Erd\"os number are close 
to the outer boundaries of the stable islands which are last reached when 
starting from the hub. 

\begin{figure}[t]
\begin{center}
\includegraphics[width=0.48\textwidth]{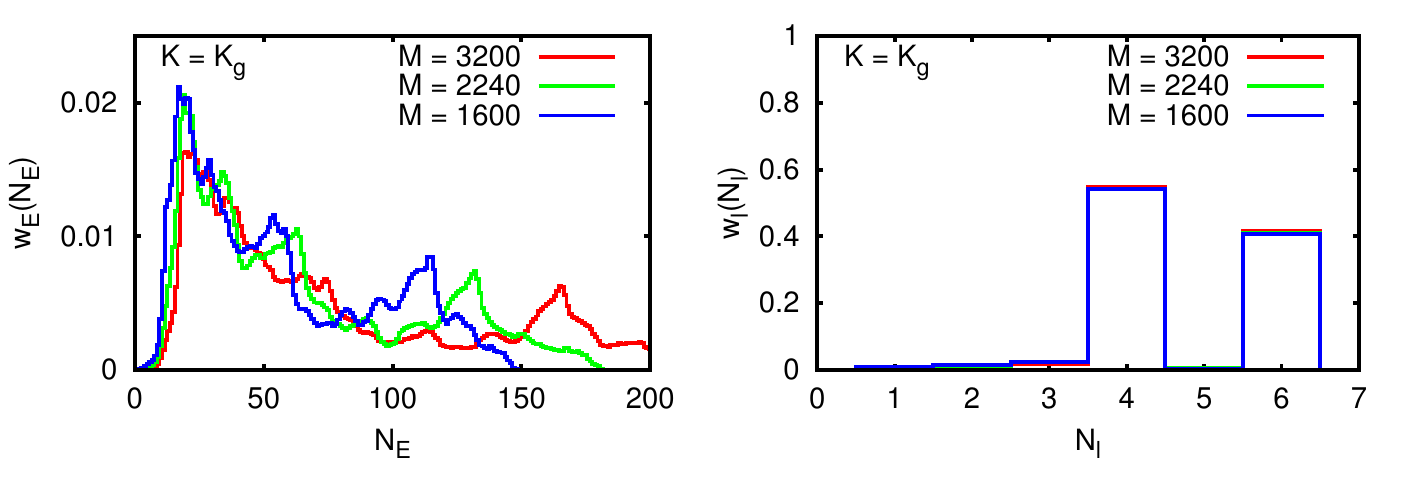}
\end{center}
\caption{\label{fig5}
Same as in Fig.~\ref{fig1}
but for the golden critical value
$K=K_g=0.971635406$ of map (\ref{eq_stmap}).
}
\end{figure}

Fig.~\ref{fig5} shows the probability distributions of $N_E$ and $N_l$ 
for the standard map at the golden critical value $K=K_g=0.971635406$ with 
a complicated structure of stable islands inside the main chaotic component. 
The distribution of $N_E$ is now rather large with non-vanishing 
probabilities at values $N_E\sim 10^2$ and several local maxima 
due to the complicated phase space structure with different layers of initial 
diffusive spreading (limited by the golden curve). 
The distribution of $N_l$ is mostly concentrated on the two values 
$N_l=4$ and $6$ in agreement with the above discussion since 
$N_l^{(\rm max)}=2\lceil 2+K_g\rceil=6$.

\begin{figure}[t]
\begin{center}
\includegraphics[width=0.48\textwidth]{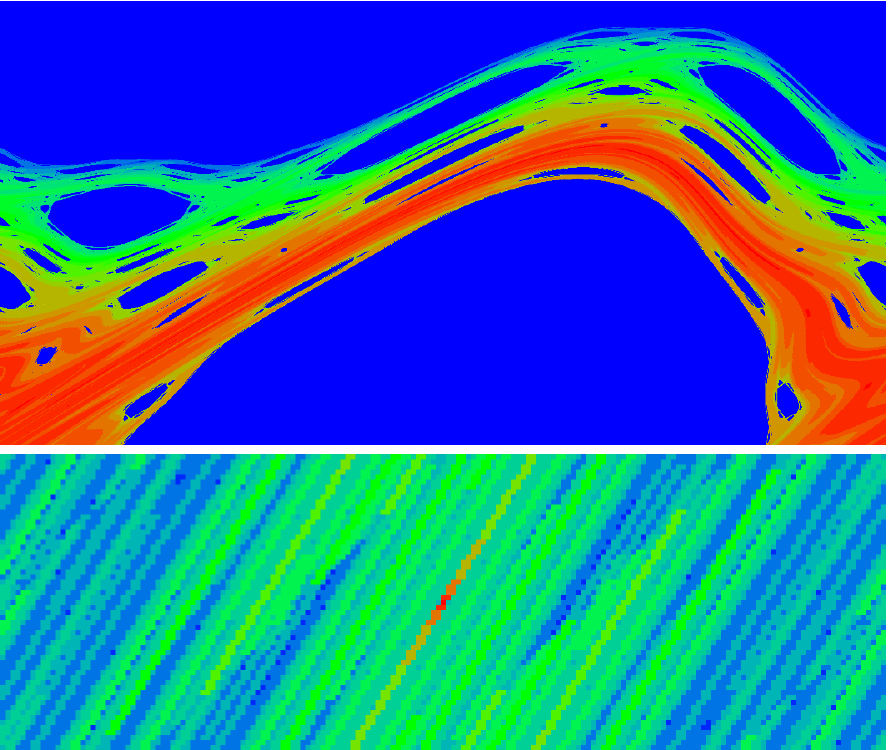}
\end{center}
\caption{\label{fig6}
Density plots of Erd\"os number (similar as in bottom panels of 
Fig.~\ref{fig4}) 
for the Ulam network of the Chirikov standard map at the golden critical value
$K=K_g=0.971635406$ for $M=3200$ (top panel, hub cell at position 
$x=0.1/(2\pi)$, $p=0$) 
and of the Arnold cat map for $M=59$, $L=3$ (bottom panel, hub cell at 
position $x=p=0$). 
In the latter case the roles of $x$- and $p$-axis have been exchanged 
for a better visibility. 
}
\end{figure}

The spatial distribution of $N_E$ for $K=K_g$ (using a hub cell at 
$x=0.1/(2\pi)$ and $p=0$) is illustrated in the top 
panel of Fig.~\ref{fig6} by the same type of color plot used 
for the lower panels of Fig.~\ref{fig5}. In this case $N_E$ follows 
clearly the (very slow) diffusive spreading with smallest $N_E$ values 
in the layers close to the hub and maximal $N_E$ values closest to 
the top layers just below the golden curve. 

The bottom panel of Fig.~\ref{fig6} shows the $N_E$ color plot (with hub cell 
at $p=x=0$) for the Arnold cat map at $L=3$ and the rather small value $M=59$ 
for a better visibility. 
As for the case $K=7$  of map (\ref{eq_stmap}) the 
Erd\"os number follows a one-dimensional unstable manifold (a straight 
refolded line for the cat map) and the chaotic spreading reaches quite quickly 
the two outer square boxes (with $|p|>0.5$). We have verified that this 
behavior is also confirmed by the corresponding $N_E$ color plots at larger 
values of $M$. The evolution of the nodes with smallest $N_E$ values does 
not follow the classical diffusion which can be understood by the fact 
that the Erd\"os number only cares about reaching a cell as such even with 
a very small probability while the diffusive dynamics applies to the 
evolution of the probability occupation of each cell. This is similar 
to a one-dimensional random walk with a diffusive spreading $\sim\sqrt{Dt}$ 
of the spatial probability distribution while the Erd\"os number (i.e. 
set of ``touched'' cells) increases ballistically in time $\sim t$.

\begin{figure}[t]
\begin{center}
\includegraphics[width=0.48\textwidth]{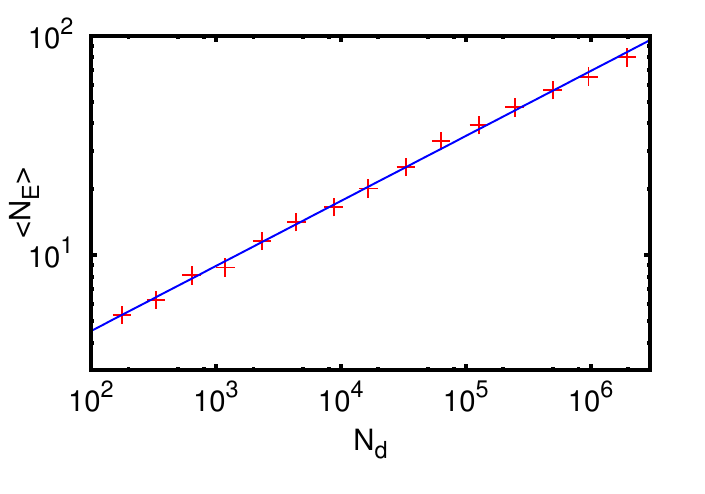}
\end{center}
\caption{\label{fig7}
Dependence of the average  Erd\"os number $\langle N_E\rangle $
on the number of nodes (cells) $N_d$ for the Ulam network
of the Chirikov standard map at $K=K_g$ (red plus symbols)
with hub cell at $x=0.1/(2\pi)$, $p=0$ in a double logarithmic 
representation. 
The blue line corresponds to the fit 
$\langle N_E\rangle=C N_d^b$ with $C=1.15\pm 0.04$ and 
$b=0.297\pm 0.004$. }
\end{figure}

Fig.~\ref{fig7} shows the dependence of the average $\langle N_E\rangle$
on $N_d$ for $K=K_g$ which follows a power law 
$\langle N_E\rangle\sim N_d^b$ with $b=0.297\pm 0.004$. 
For this case the logarithmic behavior 
$\langle N_E\rangle\sim \ln N_d$ (\ref{eq_erdos}) observed for $K\ge 5$ is not valid due to the 
small Lyapunov exponent and complicated phase space structure with slow 
diffusive spreading and complications from orbits trapped around stable 
islands. 

\begin{figure}[t]
\begin{center}
\includegraphics[width=0.48\textwidth]{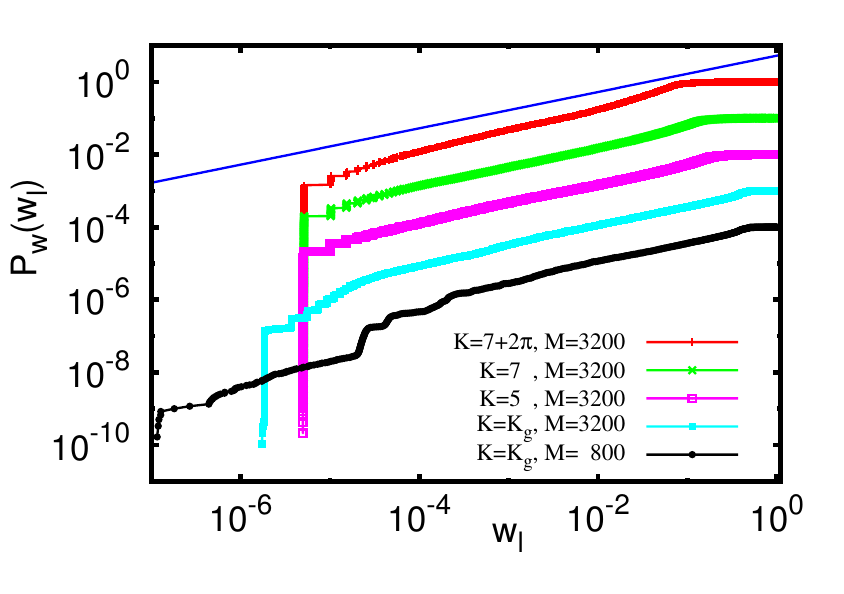}
\end{center}
\caption{\label{fig8}
Double logarithmic representation of fraction $P(w_l)$ of 
links with weight below $w_l$ of the Ulam network for the 
Chirikov standard map with various values of $K$ and $M$. The lower curves 
are successively shifted down by a factor of 10 for a better visibility. 
The straight blue line corresponds to the simplified 
power law behavior $\sim w_l^{0.5}$. 
The power law fits $P_w(w_l)=C\,w_l^b$ for the range 
$5\times 10^{-5}<w_l<2.5\times 10^{-2}$ for 
$K\ge 5$ ($10^{-4}<w_l<5\times 10^{-1}$
for $K=K_g$) provide the fit values (top to bottom curves)~:
$C=2.63\pm 0.03,\ 1.64\pm 0.01,\ 1.68\pm 0.02,\ 1.31\pm 0.01, 1.39\pm 0.02$ 
and 
$b=0.584\pm 0.002,\ 0.521\pm 0.001,\ 0.525\pm 0.002,\ 0.542\pm 0.001,\ 
0.571\pm 0.003$.
The vertical lines of data at the left side correspond to 
the smallest possible weight values $N_d/10^{12}$ being 
the typical inverse number of trajectory crossings per cell 
with $N_d\approx M^2/2$ for $K\ge 5$ or 
$N_d\approx M^2/4$ for $K=K_g$. 
}
\end{figure}

To explain the obtained dependence
$N_E \sim N_d^{0.3}$ we give the following heuristic argument. 
According to the renormalization description of the
critical golden curve the typical time scale of motion
in the vicinity of a certain resonance with the Fibonacci approximate of
the golden rotation number $r_n = q_{n-1}/q_n \rightarrow r_g =(\sqrt{5}-1)/2$
with $q_n=1,2,3,5,8,...$ is $t_n \sim q_n$ (same for the symmetric 
golden curve with $r=1-r_g$) \cite{mackay,chirikov1999}. 
At the same time the area of one cell close to the resonance $q_n$ with 
typical size $1/q_n^2$ scales approximately as 
$A_n \sim 1/(q_n^2\,t_n)\sim 1/q_n^3$. Since a cell of the Ulam network 
has an area $1/N_d \sim A_n$ we obtain that $t_n \sim N_d^{1/3}$.
We expect that the typical time to reach the resonance with 
largest $q_n$ value that can 
be resolved by the UPFO discretization is of the order of the most 
probable Erd\"os number such that $N_E \sim t_n \sim N_d^{1/3}$ leading to
$b=1/3$ comparable with the obtained numerical value. Of course, 
this handwaving argument is very simplified since in addition to 
Fibonacci resonance approximates there are other 
resonances which play a role in long time sticking of trajectories
and algebraic decay of Poincar\'e recurrences
(see e.g. \cite{ketzmerick,ulampoincare,fishman2017}). Also as discussed above
the Erd\"os number is for a network with equal weights of transitions
while in the UPFO for the Chirikov standard map the transition weights 
are different.

Indeed, 
since the Erd\"os number does not depend on the weight $w_l$ of a link it 
follows in principle a different dynamics than the UPFO applied on an initial 
localized state. Therefore we also analyzed the statistical distribution of 
link weights $w_l$. Fig.~\ref{fig8} shows the integrated weight distribution 
$P_w(w_l)$ (fraction of links with weight below $w_l$) of the UPFO for the 
Chirikov standard map for different values of $K$ and $M$. 
The vertical lines at 
some minimal value correspond to the smallest possible weight values 
$w_l^{\rm (min)}=N_d/10^{12}$ being the typical inverse number of 
trajectory crossings per cell and are due to the finite length of the 
the iteration trajectory. Apart from this, in the regime 
$w_l^{\rm (min)}<w_l<0.1$, the behavior is very close 
to a power law $P_w(w_l)\sim w_l^b$ with some exponent rather close to 
$b\approx 0.5$ depending on $K$ values and fit ranges. This leads to 
a square root singularity in the probability distribution 
$p_w(w_l)={P_w}'(w_l)\sim w_l^{-0.5}$. 

To understand this dependence we 
remind that according to the discussion of Appendix \ref{appenda} 
the weights $w_l$ are 
given as the relative intersection areas of a certain parallelogram (being 
the image of one Ulam cell by the map) with the target Ulam cells 
and that the bottom corner point of the parallelogram (relative to its 
target cell) is given by $\Delta s(\xi_0,\,\xi_0)$ where $\xi_0\in[0,1[$ 
has a uniform quasi-random distribution (see also bottom right panel of 
Fig.~\ref{fig11} in Appendix \ref{appenda}). If $1-\xi_0\ll 1$ this provides 
the triangle area (relative to the cell size $\Delta s^2$) being:
$w_l=C(A)(1-\xi_0)^2/2$ with a coefficient dependent on the parameter 
$A=K\cos(2\pi \Delta s\,x_i)$ and also if we consider the triangle in 
the cell around the lowest corner point or the cell right next to it 
(which may have a smaller area depending on $A$). 
Since $\xi_0$ is uniformly distributed we find (after an additional 
average over the initial cells, i.e. over the parameter $A$) 
immediately that $p_w(w_l)\sim w_l^{-1/2}$. It is also possible that 
the top corner point of the parallelogram (instead of the bottom corner point) 
may produce the minimal weight (among all target cells for a given initial 
cell). However, the top corner point also lies on the diagonal (relative 
to its target cell) and produces therefore the same square root singularity.

The appearance of the singularity is certainly very interesting. However, 
this singularity is integrable and the main part of 
links still have weights $w_l$ comparable to 
its typical value $w_l\sim N_l^{-1}$ given by the relative intersection 
areas of the parallelogram with the other target cells. 
Furthermore, despite this 
singularity, it seems that the dynamics of the Erd\"os number follows 
qualitatively quite well the chaotic dynamics induced by the direct 
application of the UPFO as can  be seen for example in Figs.~\ref{fig4} 
and \ref{fig6}. 

The results of this Section show that in the regime of strong chaos
the Ulam networks are characterized by small values of the Erd\"os number
$N_E \sim \ln N_d$ growing only logarithmically with the network size $N_d$.
However, the presence of stability islands can modify the asymptotic behavior
leading to a more rapid growth with $N_E \sim N_d^{0.3}$
as it is the case for the critical golden curve of the Chirikov standard map
where a half of the total measure is occupied by stability islands.

\begin{figure}[t]
\begin{center}
\includegraphics[width=0.48\textwidth]{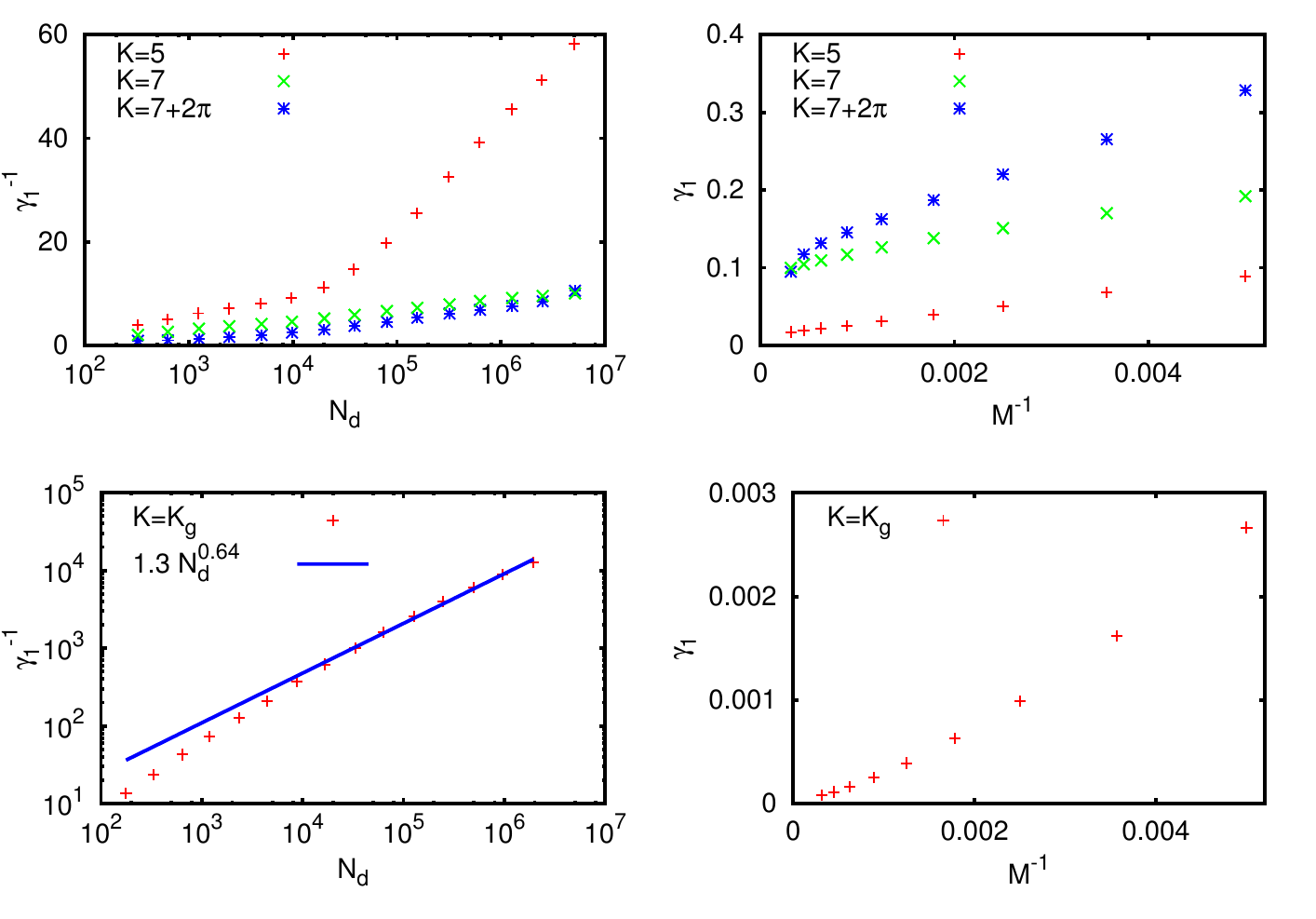}
\end{center}
\caption{\label{fig9}
Dependence of the decay rate 
$\gamma_1=-2\ln|\lambda_1|$ on network size where 
$\lambda_1$ is the second eigenvalue of the UPFO for the Chirikov standard 
map. Top panels correspond to the values $K=5,\,7,\,7+2\pi$ and bottom 
panels to $K=K_g$. Left panels show the dependence of $\gamma_1^{-1}$ 
on network size $N_d$ with a logarithmic representation for $N_d$ 
(and also for $\gamma_1^{-1}$ for bottom panel with $K=K_g$).
Right panels show the dependence of $\gamma_1$ on $M^{-1}$ where 
$N_d\approx M^2/2$ ($M^2/4$) for the cases with $K\ge 5$ ($K=K_g$). 
The blue line in bottom left panel corresponds to the power law fit: 
$\gamma_1^{-1}=C(N_d)^b$ with $C=1.3\pm 0.2$ and $b=0.64\pm 0.01$ 
for the fit range $N_d>10^4$. }
\end{figure}

\section{Small relaxation rates of UPFO}
\label{sec5}
The average (or maximal) Erd\"os number gives the time scale at which 
the UPFO touches most (or all) Ulam cells when applied to an initial state 
localized at one cell (hub) but it does not take into account 
the probability density associated to the target cells which may be very 
small for the cells with largest $N_E$ at iteration times 
$t\sim N_E^{\rm (max)}$. However, the direct iterated 
application of the UPFO on a typical localized initial state converges 
exponentially versus a (roughly) uniform stationary distribution (for 
the accessible cells) as 
$\ \sim\exp(-\gamma_1 t/2)$ where the decay rate 
is given by $\gamma_1=-2\ln(|\lambda_1|)$ in terms of the second eigenvalue 
$\lambda_1$ of the UPFO (with the first eigenvalue always being $\lambda_0=1$ 
for a non-dissipative map and its eigenvector being the stationary 
homogeneous density
distribution over the chaotic component in the phase plane). 

First results for $\gamma_1$ were given 
for the Chirikov standard map in \cite{frahmulam} and the Arnold cat map 
in \cite{arnoldcat}. Here we present new results for $\gamma_1$ 
obtained by the Arnoldi method for additional values of $K$ and larger $M$. 
In most cases an Arnoldi dimension of $n_A=1000$ (see Ref. \cite{frahmulam} 
for computational details) is largely sufficient to get numerical 
precise values of $\gamma_1$ as well as a considerable amount of largest 
complex eigenvalues. Only for the Chirikov standard map at $K=K_g$, where 
the eigenvalue density close to the complex unit circle is rather 
elevated, we used $n_A=3000$ ($4000$) for $M\le 1600$ ($1600<M\le 3200$).

Fig.~\ref{fig9} shows two different representations of the dependence 
of $\gamma_1$ on $M$ or $N_d\sim M^2$ for the standard map and our usual 
values $K=K_g,\,5,\,7,\,7+2\pi$. 
For $K\ge 5$ the plot of the top left panel seems to indicate 
that $\gamma_1^{-1}\sim C_1+C_2\ln(N_d)$ 
(with two different regimes for $K=5$) possibly indicating that 
$\ \gamma_1\sim 1/\ln(N_d)\to 0$ for very large system size. 
However, the alternative plot of $\gamma_1$ versus $1/M$ in top right panel 
might indicate a finite limit of $\gamma_1$ for $M\to\infty$ at least for 
$K=7$ with a very particular classical behavior due to the stable island 
\cite{chirikov2000} visible in (bottom left panel of) Fig.~\ref{fig4}. 
We think that the numerical data does not allow to conclude clearly 
if the infinite size limit of $\gamma_1$ is vanishing or finite since 
the possible logarithmic behavior may manifest itself at extremely 
large values of $M$ or $N_d$ numerically not accessible. 

For $K=K_g$ we confirm the power law behavior 
$\gamma_1^{-1}\approx 1.3\,N_d^{0.64}$ for $N_d>10^4$ in agreement with 
the results of \cite{frahmulam}. However, as discussed in \cite{frahmulam}, 
taking into account the data with $N_d<10^4$ one may also try a more 
complicated fit using a rational function in $1/M$ providing a different 
behavior $\gamma_1^{-1}\sim N_d^{0.5}\sim M$ but this would be visible 
only for extremely large, numerically inaccessible, values of $M$. 
Thus for the case $K=K_g$ we can safely conclude that $\gamma_1\to 0$ 
for $M\to\infty$ in agreement with the power law statistics of the 
Poincar\'e recurrence time at $K=K_g$. 

\begin{figure}[t]
\begin{center}
\includegraphics[width=0.48\textwidth]{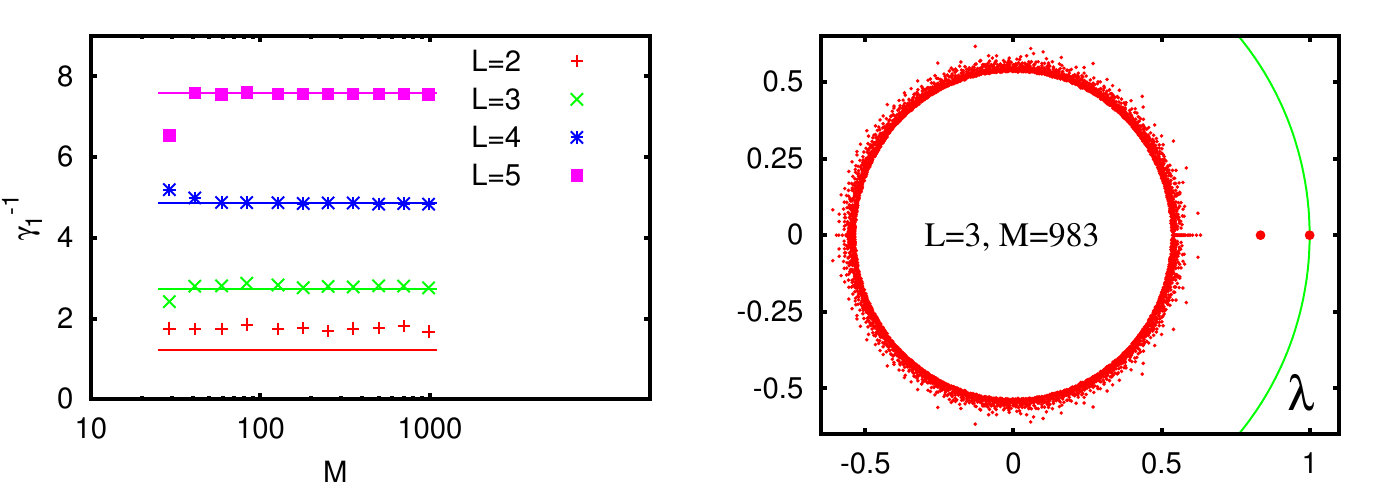}
\end{center}
\caption{\label{fig10}
{\em Left panel:} Dependence of the inverse decay rate 
$\gamma_1^{-1}=(-2\ln|\lambda_1|)^{-1}$ on $M$ where 
$\lambda_1$ is the second eigenvalue of the UPFO for the Arnold cat map
for the cases $L=2,\,3,\,4,\,5$ and certain prime numbers 
$29\le M\le 983$. The horizontal lines correspond 
to the theoretical values 
$(\gamma_1^{\rm (diff)})^{-1}=(D(2\pi)^2/L^2)^{-1}=3L^2/\pi^2$ 
based on diffusive dynamics in $p$-direction with diffusion 
constant $D=1/12$, finite system size $L$ and periodic 
boundary conditions. 
{\em Right panel:} Complex spectrum of top $\sim 4000$ reliable 
eigenvalues $\lambda_j$ of the 
UPFO for the Arnold cat map at $L=3$, $M=983$ obtained by the 
Arnoldi method with Arnoldi dimension 5000 and using a symmetrized version 
of the map in $p$-direction for $0\le p<L/2$ 
(symmetrized matrix size 
$N_d^{\rm (sym)} \approx L M^2/2 \approx 1.5 \times 10^6$). The green curve 
corresponds to the unit circle. The first two eigenvalues 
$\lambda_0=1$ and $\lambda_1\approx 0.834183$, corresponding to 
$\gamma_1\approx 0.362604$, are shown as
red dots of increased size.}
\end{figure}

Concerning the Arnold cat map the very efficient algorithm to compute the UPFO 
described in Appendix \ref{appenda} 
combined with the Arnoldi method allows to treat rather 
large values of $M$, e.g. up to $M=983$ corresponding to 
$N_d\approx 3\times 10^6$. We remind that due to the necessity to store 
simultaneously $\sim n_A$ vectors of size $N_d$ 
it is not possible to consider the Arnoldi method for values 
such as $M=14699$ for which we were able to compute the Erd\"os number
only using the network link structure. 
We find that apart from $\lambda_0=1$ (nearly) all real and complex 
eigenvalues of the UPFO 
are double degenerate due to the symmetry $p\to -p$ and $x\to -x$. Therefore 
we also implemented a symmetrized version of the UPFO for the cat map where 
cells at $p_i<0$ are identified with the corresponding cell at $p_i>0$ 
(and $x_i\to -x_i$). 
This allows the reduction of $N_d$ by roughly a factor of two (cells at 
$p_i=0$ are kept as such) and lifts the degeneracy allowing to obtain more 
{\em different} eigenvalues at given value of $n_A$. For small values of $M$ 
the symmetrized version may miss a few eigenvalues but at $M=983$ we find 
that the spectra coincide numerically (for the amount of reliable eigenvalues 
which we were able to compute for the non-symmetrized UPFO).
Concerning the computation of $\gamma_1$ this point is not important since 
$n_A=100$ is already sufficient (both symmetrized and non-symmetrized UPFO) 
but we verified all $\gamma_1$ values also with $n_A=1000$. 
However, the symmetrized UPFO allows to obtain larger spectra with less 
effort. 

The left panel of Fig.~\ref{fig10} shows the dependence of 
$\gamma_1^{-1}$ on $M$ for $2\le L\le 5$ and we see for all cases 
a clear convergence for $M\to\infty$ with final constant values 
already for $M>50$. The numerical values of $\gamma_1$ 
can be explained by the diffusive dynamics (\ref{cateq2}) in 
$p$-direction with $D=1/12$ which is expected to be valid for sufficiently 
large $L$. Assuming periodic boundary conditions in $p$-direction 
the diffusive dynamics implies the classical 
theoretical decay rate $\gamma_1^{\rm (diff)}=D(2\pi)^2/L^2=\pi^2/(3L^2)$ 
which agrees quite accurately with our numerical values for $L\ge 3$
(this was also seen in \cite{arnoldcat} for smaller $M$ values). 
Only for $L=2$ the numerical value of $\gamma_1^{-1}$ is roughly a third 
larger than the theoretical value which is not astonishing due 
to the modest value of $L=2$ limiting the applicability of the diffusive model.

Furthermore, we show as illustration in the right panel of Fig.~\ref{fig10} 
the top spectrum 
of $\sim 4000$ eigenvalues for the case $M=983$, $L=3$ obtained by 
the Arnoldi method for $n_A=5000$ applied to the symmetrized UPFO. We 
note that apart from both top eigenvalues ($\lambda_0=1$ and 
$\lambda_1\approx 0.834183$) the spectrum is limited 
to a complex circle of radius $\approx 0.6$ with a quite particular 
pattern for the top eigenvalues with $0.55<|\lambda_j|<0.6$ and 
a cloud of lower eigenvalues with $|\lambda_j|<0.55$. 

The results of Fig.~\ref{fig10} for the Arnold cat map
clearly show that the Erd\"os number, shown in Fig.~\ref{fig3}, 
is not directly related
with the relaxation time $1/\gamma_1$ of the UPFO. 
As already discussed above on an example of a diffusive process this is 
related to the fact that the Erd\"os number does not take into account 
the variations of transition weights and measures the time when a 
cell is first touched, leading to a ballistic type of propagation 
instead of diffusion, while the relaxation time measures the convergence 
to the stationary homogeneous probability distribution for long time scales.

\section{Discussion}
\label{sec6}
We analyzed the properties of Ulam networks 
generated by dynamical symplectic maps. Our results show that 
in the case of strongly chaotic dynamics these networks
belong to the class of small world networks 
with the number of degrees of separation,
or the Erd\"os number $N_E$, growing logarithmically 
with the network size $N_d$. This growth is related 
to the Lyapunov exponent of chaotic dynamics.
However, the obtained results show that in presence of
significant stability islands the Erd\"os number growth is 
stronger with $N_E \sim N_d^{0.3}$ being related to orbits sticking
in a vicinity of islands. We also show that the Erd\"os number
is not directly related to the largest relaxation times
which remain size independent in the case of a diffusive process
like for the Arnold cat map on a long torus. 
We hope that our results will stimulate further useful inter-exchange 
between the fields of dynamical systems and directed complex networks.

\vspace{1cm}
\textit{Acknowledgments.--} This work was supported in 
part by the Programme Investissements
d'Avenir ANR-11-IDEX-0002-02, 
reference ANR-10-LABX-0037-NEXT (project THETRACOM).
This work was granted access to the HPC resources of 
CALMIP (Toulouse) under the allocation 2018-P0110. 

\appendix 
\section{Exact UPFO for the Arnold cat map}
\label{appenda}

\begin{figure}[t]
\begin{center}
\includegraphics[width=0.48\textwidth]{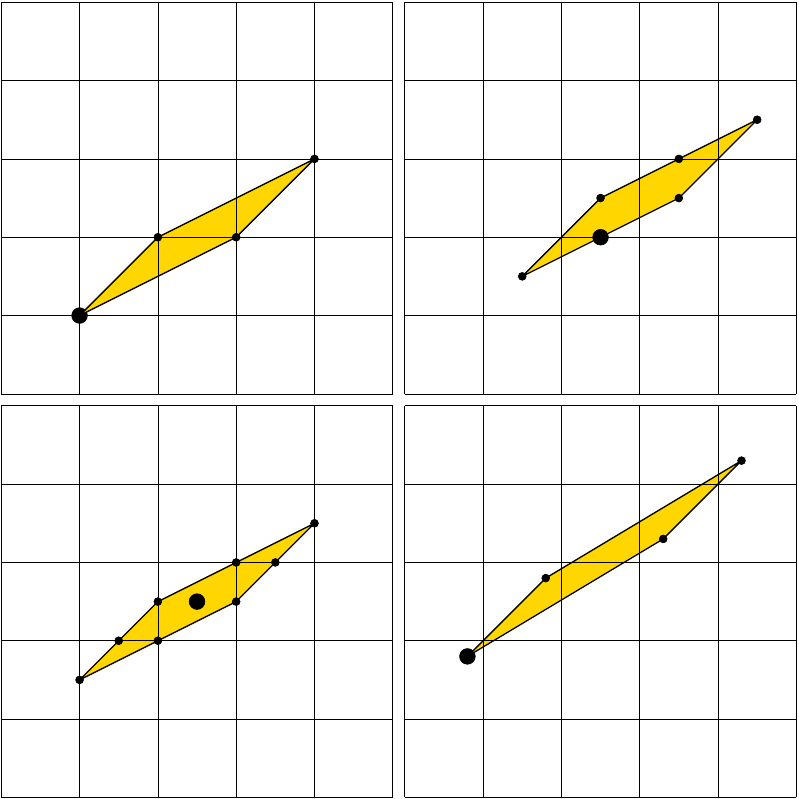}
\end{center}
\caption{\label{fig11}
Parallelogram image of one initial square Ulam cell 
shown in the grid of possible 
target cells. The panels correspond to the Arnold cat map with both 
$M$ and $LM$ even (top left panel), $M$ odd and $LM$ even (top right panel), 
both $M$ and $LM$ odd (bottom left panel) and the Chirikov standard 
map with parameters $\xi_0=0.8$ and $A=1.5$ (see text; bottom right panel). 
The thick black dots correspond to the map image of a reference 
point on the grid $(x_i\Delta s,\,p_i\Delta s)$ with integer $x_i$, $p_i$ and 
linear cell size $\Delta s=1/M$. The relative intersection areas of the 
parallelogram with possible target cells 
provide exactly (approximately) the transition probability of the UPFO 
for the cat (standard) map. For the Arnold cat map we have 
four target cells with probability $1/4$ (both $M$ and $LM$ even), 
two target cells with probability $3/8$ 
and four target cells with probability $1/16$ ($M$ odd and $LM$ even) or 
one target cell with probability $1/2$ and four target cells with 
probability $1/8$ (both $M$ and $LM$ odd).
}
\end{figure}

The exact linear form of (\ref{cateq1}) allows to compute {\em exactly} 
(without any averaging procedure) the transition probabilities needed for the 
UPFO of the Arnold cat map. 
For this we write each phase space point in the form 
$x=x_i\,\Delta s+\Delta x$ and $p=p_i\,\Delta s+\Delta p$ with $\Delta s=1/M$ 
being the linear cell 
size and $x_i$, $p_i$ being integer values. Depending on the parity of 
$M$ (or $LM$) we have $0\le \Delta x<\Delta s$ ($0\le \Delta p<\Delta s$) 
for even $M$ (even $LM$) or $-\Delta s/2\le \Delta x<\Delta s/2$ 
($-\Delta s/2\le \Delta p<\Delta s/2$) 
for odd $M$ (odd $LM$) such that each value of the integer vector 
$(x_i,p_i)$ corresponds exactly to one Ulam cell. The image of the grid 
point $\Delta s(x_i,p_i)$ by the cat map is exactly another grid point 
$\Delta s(\bar x_i,\bar p_i)$ with integer values $\bar x_i$ and $\bar p_i$. 
These grid points are either at the left (bottom) corner/boundary of the 
corresponding Ulam cell for even values of $M$ (or $ML$) or in the middle 
of the Ulam cell for odd values of $M$ (or $ML$).

The image of an initial Ulam square cell under the Arnold cat map becomes 
a parallelogram of the same area, spanned by the two vectors 
$(\Delta s,\,\Delta s)$ and $(2\Delta s,\,\Delta s)$, 
which intersects with $4$ (both $M$ and $LM$ even), $5$ 
(both $M$ and $LM$ odd) or $6$ target cells ($M$ odd but $LM$ even) as can 
be seen in Fig.~\ref{fig11}. The relative 
intersection areas of the parallelogram with each cell provide the exact 
theoretical transition probabilities given as multiples of small powers of 
$1/2$. 
For example for the most relevant case of this work, where both $M$ and 
$LM$ are odd, there are for each initial cell one target cell with transition 
probability of $1/2$ and four other target cells with probability $1/8$. 
For the other cases we have four target cells with probability $1/4$ 
(both $M$ and $LM$ even) or two target cells with probability $3/8$ 
and four target cells with probability $1/16$ ($M$ odd and $LM$ even). 

Furthermore, Fig.~\ref{fig11} also shows the relative positions of the 
concerned target cell with respect to a reference point being the image of 
the grid point of the initial Ulam cell. 
In this way it is possible to compute very efficiently and 
directly the {\em exact} Ulam network for the Arnold cat map which allowed 
us to choose $M$ up to $M=14699$ corresponding to the network size 
$N_d=LM^2\approx 6.5 \times 10^8$. We have also verified that our exact 
computation scheme is in agreement with the two other variants of the Ulam 
method (apart from statistical fluctuations in the latter).

We may also try a similar analysis of the UPFO for the Chirikov standard map 
which gives 
three complications: (i) the standard map is only locally linear for large 
values of $M$ and the scheme will only be approximate due to non-linear 
corrections; (ii) we have to add a certain (rather random/complicated) 
offset $\xi_0\Delta s$ (with 
$\xi_0=K\sin(2\pi x_i\Delta s)/(2\pi\Delta s)\mod 1$)
in the above expressions in terms of $x_i$ or $p_i$ since an initial point 
on the integer grid is no longer exactly 
mapped to another point of this grid as it was the case with the Arnold 
cat map, 
and finally (iii) the parallelogram is now spanned by the two vectors 
$\Delta s(1,1)$ and $\Delta s(1+A,A)$. Here the parameter 
$A\approx K\cos(2\pi \Delta s\,x_i)$
may take rather large values depending on $K$ and depends on the phase space 
position $x\approx \Delta s\,x_i$. The bottom right panel of Fig.~\ref{fig11} 
shows an example of such a shifted parallelogram with $\xi_0=0.8$ and 
$A=1.5$. 

For these reasons, this scheme is not suitable to construct numerically 
a reliable UPFO for the map (\ref{eq_stmap}). However, it is 
still very useful to understand quite well the distribution of the 
number $N_l$ of connected cells from one initial cell and also 
the square root singularity in the distribution of weights $p_w(w_l)$ 
of the UFPO for the standard map (see discussions in Section \ref{sec4} for 
both points).


\end{document}